\documentclass{WileyMSP-template}

\usepackage{xcolor}
\usepackage{amsmath}

\begin{document}

\pagestyle{fancy}
\rhead{\includegraphics[width=2.5cm]{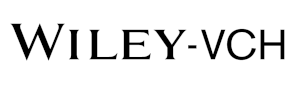}}

\title{Breaking the Exponential: Decoherence-Driven Power-Law Spontaneous Emission in Waveguide Quantum Electrodynamics}

\maketitle


\author{Stefano Longhi}



\begin{affiliations}
S. Longhi\\
Address1: Dipartimento di Fisica, Politecnico di Milano, Piazza L. da Vinci 32, I-20133 Milano, Italy\\
Address2: IFISC (UIB-CSIC), Instituto de Fisica Interdisciplinar y Sistemas Complejos - Palma de Mallorca, Spain\\

Email Address: stefano.longhi@polimi.it
\end{affiliations}


\begin{abstract}
We investigate the spontaneous emission of a two-level system coupled to a photonic waveguide, showing that dynamical dephasing in the photon modes profoundly alters the decay law. In the absence of dephasing, the emitter displays conventional exponential decay followed by a long-time power-law tail -- observable only at extremely low survival probabilities. Strikingly, when dephasing is introduced, a robust power-law decay emerges already at short times, driven by photon diffusion in the dynamically disordered environment rather than spectral edge effects. These results reveal a novel, decoherence-induced mechanism for non-exponential spontaneous emission in waveguide QED platforms.
\end{abstract}


\section{Introduction}
Exponential decay is a ubiquitous phenomenon across many areas of physics, including high-energy physics, nuclear physics, and quantum electrodynamics.  Foundational quantum-mechanical descriptions of exponential decay date back to Gamov's model of $\alpha$-decay, which introduced the concept of quantum tunneling in nuclear disintegration \cite{D-1}, and to the Weisskopf-Wigner theory of spontaneous emission of photons, developed in the 1930s to describe the radiative decay of excited atomic states\cite{D0}.
Despite their success, such models exhibit a fundamental limitation: they often rely on Hamiltonians that lack a true ground state. It has been argued that purely exponential decay requires the Hamiltonian to be unbounded both from below and above -- an unphysical condition, as real systems must possess a lowest energy state to ensure stability. As a result, purely exponential decay is widely regarded as an idealization rather than an exact physical behavior (see, e.g., \cite{D1,D2,D3,D4} and references therein). 
In reality, spontaneous emission and other quantum decay processes are known to deviate from an exact exponential law at both short and long timescales \cite{D1,D2,D3,D4,D5,D6,D7,D7b,D8,D9,D10,D10b,D10c,D10cbis,D10d,D10e,D10f}, or even at any time scale in certain models \cite{D10g,D10h}.  At very short times, the survival probability of the excited state exhibits a quadratic behavior known as the Zeno regime, which underlies the quantum Zeno and anti-Zeno effects observed under repeated measurements \cite{D11,D12,D13,D14,D15,D16,D17,D18,D19,D20,D21}. At long times, the decay transitions to a power-law behavior arising from the boundedness from below of the continuum spectrum, or more generally, from spectral edge effects \cite{D2,D3,D4,D5,D8,D9,D10,D10b,D10c,D21b}. This long-time algebraic decay can be understood through the Paley-Wiener theorem, which constrains the Fourier transform of functions with lower-bounded spectral support, making power-law tails a universal feature of physical decay processes \cite{D22,D22b}.
Beyond these universal features, it is well established that engineering the density of states of the continuum, the presence of disorder or bound states in the decay environment, or considering  giant atoms rather than point-like emitters can induce strong non-Markovian effects, substantially modifying the decay dynamics. Such structured or disordered continua as well as giant atoms may lead to pronounced deviations from exponential decay, including revivals, trapping, and long-lived coherence effects (see e.g. \cite{D23-1,D23,D23b,D23c,D23d,D23e,D23ebis,D23f,D23g}). Moreover, decoherence itself can generally introduce deviations from exponential decay laws, as demonstrated in recent studies \cite{D24}. However, power-law decay tails remain notoriously elusive in experiments \cite{D24b,D25,D25b,D25c,D25cbis,D25d,D25e}, as they manifest only after the survival probability has diminished to very small values during the preceding exponential regime. As a result, experimental evidence of long-time power-law decay is scarce \cite{D26,D27} and often relies on controllable classical photonic platforms that emulate quantum mechanical decay \cite{D27,D28,D29}.

In this work, we predict that spontaneous emission in waveguide quantum electrodynamics (QED) can exhibit power-law decay at experimentally accessible timescales when dephasing affects the electromagnetic modes of the environment. This behavior contrasts sharply with the standard exponential decay and emerges independently of spectral edge effects.
To illustrate this phenomenon, we analyze a widely studied waveguide QED model  \cite{D29b,D29c,D30,D30a,D30b,D30c,D31,D32,D32b,D33,D33b,D34,D35,D36}  consisting of a coupled resonator optical waveguide, where a two-level quantum emitter is embedded in one of the resonators, and each cavity mode is subject to local dephasing noise \cite{D36b,D36c}. We show that the resulting dynamically disordered photonic environment leads to photon diffusion, which in turn gives rise to a robust power-law decay of the emitter's population on timescales much shorter than those typically associated with long-time algebraic tails arising from edge effects. These findings demonstrate how decoherence can fundamentally alter spontaneous emission dynamics in structured photonic systems, uncovering a new, non-Markovian mechanism for non-exponential decay in waveguide QED.

\section{Waveguide QED Model with Stochastic Dephasing}
We consider a well-studied model of spontaneous emission in waveguide QED, where a point-like two-level atom interacts with the quantized photonic modes of a semi-infinite array of coupled optical cavities (CROW) \cite{D31,D32,D32b,D33,D33b,D34,D35,D36}, as schematically illustrated in Fig.1(a). This model is relevant across a range of fields \cite{D29c,D30}, including quantum nanophotonics \cite{D23-1,D29b,D37,D38,D39} and superconducting quantum circuits \cite{D30b,D30c,D33,D36b,D40}. Additionally, we assume that the field modes in the CROW experience local pure dephasing modeled as stochastic fluctuations of the cavity resonance frequencies \cite{D36c}. Let us indicate by $|e\rangle$ and $|g \rangle$ the excited and ground states of the atom, with transition frequency $\omega_0$, placed inside the edge resonator $n=0$ of the semi-infinite array [Fig.1(a)], and by $\omega_n=\omega_c+ \delta \omega_n(t)$ the resonance frequency of the e.m. mode in the $n$-th resonator of the array, where $\omega_c \simeq \omega_0$ is the unperturbed cavity resonance frequency, equal for all the resonators, and $\delta \omega_n(t)$ are stochastic (fluctuating)  contributions with zero mean modulating the reference frequency $\omega_c$. 
 The full atom-photon Hamiltonian can be written as $H=H_0+H_{\rm noise}(t)$, where $H_0$ is the atom-photon Hamiltonian in the absence of the cavity resonance fluctuations, i.e. for $\delta \omega_n(t)=0$, and $H_{\rm noise}(t)$ accounts for the stochastic fluctuations of the cavity resonance frequencies. In the rotating-wave approximation, the Hamiltonian $H_0$ reads
\begin{eqnarray}
H_0 & = & \omega_0 |e \rangle \langle e|+ \sum_{n=0}^{\infty}  \left\{\omega_c a^{\dag}_n a_n-J(a^{\dag}_{n+1}a_n+{\rm H.c.}) \right\} \nonumber \\
&+ &g_0 \left( a_0^{\dag} |g \rangle \langle e| +{\rm H.c.} \right)
\end{eqnarray}
where $J$ is the photon hopping rate between adjacent resonators, $a^{\dag}_n$ ($a_n$) is the creation (destruction) operator of the photon field in the $n$-th resonator of the array ($n=0,1,2,3,...$), and $g_0$ is the atom-photon coupling constant. The additional noise Hamiltonian $H_{\rm noise}(t)$, yielding pure dephasing effects, reads
\begin{equation}
H_{\rm noise}(t)  =  \sum_{n=0}^{\infty}   \delta \omega_n (t) a^{\dag}_n a_n.
\end{equation}
We assume that $\delta \omega_n(t)$ are independent $\delta$-correlated Gaussian white noise processes
 with zero mean and with same variance $\gamma$, i.e. 
\begin{equation}
\overline{\delta \omega_n(t)}=0 \; ,\;\; \overline{\delta \omega_n(t) \delta \omega_m(t')}= \gamma \delta_{n,m} \delta(t-t').
\end{equation}  
where the overline denotes statistical (ensemble) average.  The quantum evolution of the atom-photon system in the presence of the noise (dephasing) term is described by a Lindblad master equation, which can be derived from the stochastic Schr\"odinger equation using standard methods \cite{D36c,D41,D42,D42b,D42c}.
Let us indicate by $| \psi(t) \rangle$ the atom-photon quantum state and by $\rho_{\rm st}(t)=| \psi(t) \rangle \langle \psi(t)|$ the stochastic density matrix, which evolves under a specific realization of the noise. Over a time interval $dt$, the evolution of  $\rho_{\rm st}(t)$ reads $\rho_{\rm st}(t+dt)=U(t) \rho_{\rm st}(t) U^{\dag}(t)$, where
\begin{equation}
U(t)=\exp(-i H_0 dt-i \sum_n a^{\dag}_n  a_n dW_n)
\end{equation}
describes the unitary evolution of the system in the time interval $(t,t+dt)$,
and $dW_n(t)=\int_{t}^{t+dt} dt' \delta \omega_n(t')$ are independent Wiener processes,
\begin{equation}
    \overline{dW_n(t)} = 0\; , \;\;  \overline{dW_n(t) dW_m(t)}= \gamma \delta_{n,m} dt.
\end{equation}
The evolution equation of the density matrix $\rho(t)=\overline {\rho_{\rm st}(t)}$ is obtained by expanding $U(t)$ up to first order in $dt$, thus including quadratic terms in $dW_n$, and taking the stochastic average \cite{D42b}. This yields the master equation in Lindblad form
\begin{eqnarray}
    \frac{d\rho}{dt} & = & \overline{ \frac{d\rho_{\text{st}}}{dt} }=-i[\hat{H}_{0}, \rho] + \\
    & + &  \gamma \sum_n \left( a^{\dag}_n a_n \rho a^{\dag}_n a_n- \frac{1}{2} \rho a^{\dag}_n a_n a^{\dag}_n a_n -\frac{1}{2} a^{\dag}_n a_n a^{\dag}_n a_n \rho \right)  \nonumber
\end{eqnarray}
where the dissipative terms in the equation account for pure dephasing effects of the photon field with a rate $\gamma$.
\textcolor{black}{We note that in realistic implementations, the dephasing noise may exhibit finite temporal or spatial correlations rather than being fully delta-correlated in time or completely uncorrelated between cavities.  However, as long as the correlation time of the noise is shorter than the characteristic photon hopping timescale in the coupled cavity system, the qualitative features of the decay dynamics remain robust. This implies that the main predictions of the model are not significantly affected by realistic finite correlations, providing guidance for potential experimental implementations. To this regard, we mention that the master equation with pure dephasing given by Eq.(6) can be obtained beyond the $\delta$-correlation limit considered here, i.e. one could assume more generally a finite correlation time for the cavity resonance frequency fluctuations $\delta \omega_n(t)$. In this case, the dephasing rate $\gamma$ entering in Eq.(6) is determined by the low-frequency component of the noise spectral density (for technical details see \cite{D36b,D36c}).  Such correlations effectively reduce the dephasing rate $\gamma$, however they do not affect the qualitative relaxation dynamics.} Finally, it should be mentioned that in the limiting case $J=0$, i.e. when the two-level atom interacts with the single bosonic mode of the resonator $n=0$, the above model reduces to the Jaynes-Cumming model with pure dephasing.

\section{Spontaneous emission decay}
In this section we investigate the dynamics of the spontaneous emission process and the influence of dephasing in such a decay process. At initial time, we assume that the atom is in the excited state with the photon field in the vacuum state $|0 \rangle$, i.e. we assume $| \psi(t=0) \rangle= |e \rangle \otimes |0 \rangle$. Since the total excitation number is conserved in the spontaneous decay dynamics, we can restrict the analysis considering the single excitation sector of Hilbert space, which is spanned by the kets 
$|e \rangle \otimes |0 \rangle$ and  $|g \rangle \otimes |n \rangle$ ($n=0, 1, 2, 3, ...$), where $|n \rangle \equiv a_n^{\dag} |0 \rangle$ corresponds to the photon state with a  single photon in the $n$-th resonator of the array. Introducing the density matrix elements 
\begin{eqnarray}
\rho_{n,m}(t) & = & \langle g| \otimes \langle n| \rho(t) |g \rangle \otimes | m \rangle =\rho_{m,n}^*(t)\\ 
\rho_{n,\epsilon}(t) & = & \langle g| \otimes \langle n| \rho(t) |e \rangle \otimes | 0 \rangle = \rho_{\epsilon,n}^*(t) \\
\rho_{\epsilon,\epsilon} (t) & = & \langle e| \otimes \langle 0 | \rho(t) | e \rangle \otimes |0 \rangle
\end{eqnarray}
from the quantum master equation (6) one obtains the following evolution equations for the density matrix elements
\begin{eqnarray}
\frac{d \rho_{n,m}}{dt} & = & i J \left( \rho_{n+1.m}+\rho_{n-1,m}-\rho_{n,m-1}-\rho_{n,m+1} \right)  \nonumber \\
& - & \gamma (1-\delta_{n,m}) \rho_{n,m}+ig_0 \left(  \delta_{m,0} \rho_{n, \epsilon}- \delta_{n,0} \rho_{\epsilon,n} \right) \\
\frac{d \rho_{n,\epsilon}}{dt} & = & \left( i \omega_0 -i \omega_c -\frac{\gamma}{2} \right) \rho_{n,\epsilon} +iJ (\rho_{n-1,\epsilon}+\rho_{n+1,\epsilon}) \nonumber \\
& -& i g_0 \delta_{n,0} \rho_{\epsilon.\epsilon}+ig_0 \rho_{n,0} \\
\frac{d \rho_{\epsilon,\epsilon}}{dt} & = & ig_0( \rho_{\epsilon,0}-\rho_{0,\epsilon})
\end{eqnarray}
which should be integrated with the initial conditions $\rho_{\epsilon,\epsilon}(0)=1$ and $\rho_{n,m}(0)=\rho_{n,\epsilon}(0)=0$.
The decay law (survival probability) describing the spontaneous emission of photon from the atom is given by
\begin{equation}
P_s(t)=\rho_{\epsilon,\epsilon}(t).
\end{equation}

 \begin{figure}
\includegraphics[width=8.5 cm]{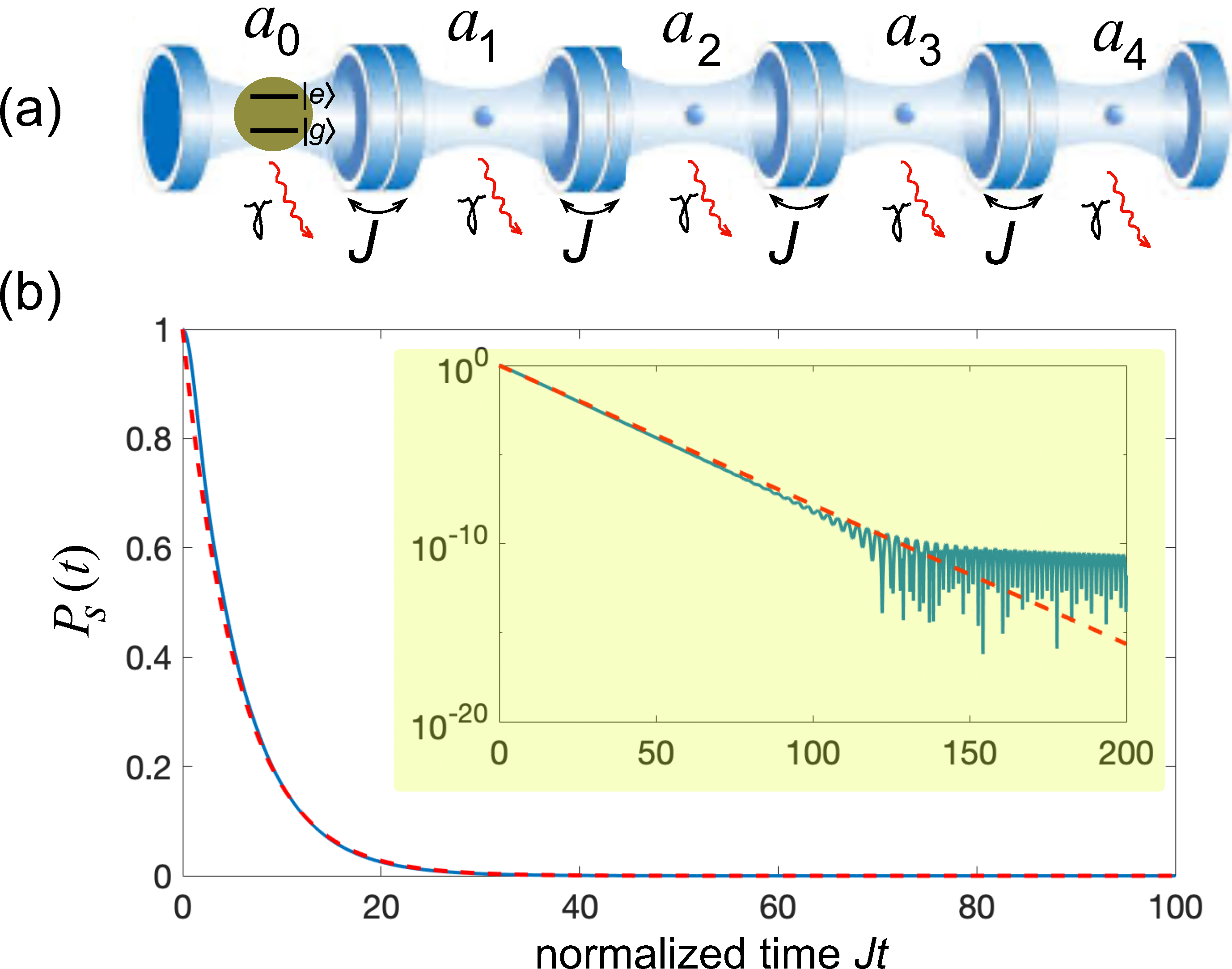}
\caption{(a) Schematic of a semi-infinite array of coupled optical cavities with a two level atom placed inside the edge chain resonator (index $n=0$). $J$ is the hopping rate of photons between adjacent resonators of the array, $\omega_0$ is the atomic resonance frequency, and $g_0$ the atom-photon coupling rate.  Dephasing of the cavity photon modes with a rate $\gamma$ is introduced by fluctuations of the cavity resonance frequencies $\omega_n(t)=\omega_c+ \delta \omega_n(t)$ around a mean value $\omega_c$. (b) Numerically-computed behavior of the survival probability of the atom, $P_s(t)$, in the spontaneous decay process in the absence of dephasing (solid blue curve), and approximate exponential decay law (dashed red curve) with a decay rate $\Gamma= 2 g_0^2/J$ as given by the Fermi golden rule. Parameter values are $\lambda=g_0/J=0.3$ and $\omega_0=\omega_c$. The inset in (b) depicts the decay curves on  a logarithmic vertical scale, clearly showing long-time deviations from an exponential decay.}
\end{figure}

In the following analysis, we assume the exact resonance $\omega_0=\omega_c$ between the atomic transition frequency $\omega_0$ and the mean cavity resonances $\omega_c$, and the weak coupling regime $g_0 \ll J$. As discussed in subsection A, in the absence of dephasing ($\gamma=0$) the emitter displays conventional exponential decay followed by a long-time power-law tail, which is observable only at extremely low survival probabilities. The long-time decay tail displays a $\sim 1/t^3$ power-law decay due to spectral edge effects of the tight-binding CROW band, with a superimposed oscillation arising from interference effects.
Strikingly, as shown in subsection C when strong dephasing is introduced, a robust power-law decay $ \sim 1/ \sqrt{t}$ emerges already at short times, driven by photon diffusion in the dynamically disordered environment rather than spectral edge effects.
 
\subsection{Spontaneous emission decay without dephasing: long-time power-law tails arising from spectral edges}
In the absence of dephasing of the photon field ($\gamma=0$), the process of spontaneous emission is exactly solvable. The system evolves remaining in a pure state, namely one has 
\begin{equation}
| \psi(t) \rangle= c_a(t) |e \rangle \otimes |0 \rangle+ \sum_n C_n(t) |g \rangle \otimes |n \rangle
\end{equation} 
with $c_a(0)=1$ and $C_n(0)=0$. The survival probability is simply given by $P_s(t)=|c_a(t)|^2$. 
The {\em exact} temporal evolution of the amplitude probability $c_a(t)$ can be written as a contour integral using standard spectral methods (see e.g. \cite{D2,D10cbis,D15,D27}), and  reads \cite{D27}
\begin{equation}
c_a(t)=\exp(-i \omega_0 t) \int_{-\infty+i0^+}^{\infty+i0^+}dE G(E)  \exp(-iEt) 
\end{equation}
where
\begin{equation}
G(E)=  \frac{1}{E(1-\lambda^2/2) +(\lambda^2/2) {\sqrt{E^2-4J^2} }}
\end{equation}
is the propagator and $\lambda=g_0/J \ll 1$ is the atom-photon coupling  rate $g_0$ normalized to the photon hopping rate $J$. For $\lambda \ll 1$,  the survival probability, $P_s(t)=|c_a(t)|^2$, is very well fitted by the exponential decay law $P_{\rm exp}(t)=\exp(- \Gamma t)$ with the decay rate $\Gamma= 2 g_0^2/J$ as given by the Fermi golden rule. However, deviations from the exponential law occurs at short times, $t < \sim \tau_1$, where the decay law is parabolic (Zeno regime), and at long times $t>  \sim \tau_2$, where the probability $P_s(t)$ shows the power-law tail $ \sim 1/t^3$ with a superimposed oscillation arising from spectral edges \cite{D27}. An illustrative example of the numerically-computed decay law is depicted in Fig.1(b). An estimate of the times $\tau_1$ and $\tau_2$ was derived in Ref.\cite{D27} and, for the resonance condition $\omega_c=\omega_0$ and for $\lambda \ll 1$, read 
\begin{equation}
\tau_1 \simeq \frac{1}{J} \; , \; \; \tau_2 \simeq \frac{1}{{\Gamma}} {\rm ln} \left( \frac{2\pi}{\lambda^{10}}\right).
\end{equation}
Note that, for $\lambda \ll 1$, the time scale $\tau_2$ becomes extremely large, so that deviations from the exponential decay law emerge only at very long times, when the survival probability $P_s(\tau_2) \simeq \exp(-\Gamma \tau_2) \sim (\lambda^{10})/(2 \pi)$ is already extremely small, as illustrated in the inset of Fig.1(b). As a result, observing the power-law decay tails becomes notoriously difficult in experiments operating in the weak coupling regime and away from spectral edges \cite{D24b,D25,D25b,D25c,D25cbis,D25d,D25e}.

\subsection{Damped vacuum Rabi oscillations}
The other simple and exactly-solvable model is provided by the limiting case $J=0$, which entails the interaction of the atom with the single quantized e.m. mode of the resonator $n=0$. In the absence of dephasing $(\gamma=0$), one recovers the usual Jaynes-Cummings model \cite{Gerry}, which in the single excitation sector of Hilbert space displays undamped vacuum Rabi oscillations at the frequency $\Omega_R=2 g_0$. When dephasing in the photon field is introduced, coherence in the dynamics is lost and the vacuum Rabi oscillations become damped and the survival probability $P_s(t)$ converges toward $1/2$, as illustrated in Fig.2(a). The relevant dynamics is captured by Eqs.(10-12) with $J=0$, yielding the following linear set of coupled equations for the excited atom population $\rho_{\epsilon,\epsilon}(t)$, the photon population $\rho_{0,0}(t)$ and the atom-photon coherences $\rho_{\epsilon.0}(t)=\rho_{0,\epsilon}^*(t)$ 
\begin{eqnarray}
\frac{d \rho_{\epsilon,\epsilon}}{dt} & = & ig_0 \left( \rho_{\epsilon,0}- \rho_{0,\epsilon }\right)  \\
\frac{d \rho_{0,0}}{dt} & = & - ig_0 \left( \rho_{\epsilon,0}- \rho_{0,\epsilon }\right)  \\
\frac{d \rho_{0,\epsilon}}{dt} & = & -\frac{\gamma}{2} \rho_{0,\epsilon}+ ig_0 \left( \rho_{0,0}- \rho_{\epsilon,\epsilon }\right) \\
\frac{d \rho_{\epsilon,0}}{dt} & = & -\frac{\gamma}{2} \rho_{\epsilon,0}- ig_0 \left( \rho_{0,0}- \rho_{\epsilon,\epsilon }\right) 
\end{eqnarray}
where we assumed $\omega_c=\omega_0$. For the initial pure state $| \psi(t=0) \rangle= | e \rangle \otimes |0 \rangle$, the above equations should be integrated with the initial conditions $\rho_{\epsilon.\epsilon}(0)=1$ and $\rho_{0,0}(0)=\rho_{0,\epsilon}(0)=\rho_{\epsilon,0}(0)=0$. The solution is attracted toward the  stationary (equilibrium) solution $\rho_{\epsilon,\epsilon}^{(s)}=\rho_{0,0}^{(s)}=1/2$ and $\rho_{0,\epsilon}^{(s)}=\rho_{\epsilon,0}^{(s)}=0$, which is the eigenstate of Eqs.(18-21) corresponding to zero eigenvalue. Typical relaxation dynamics for increasing values of the dephasing rate $\gamma$ is depicted in Fig.2(a). The relaxation toward equilibrium depends sensitively on the ratio $\gamma / g_0$, with a transition from under-damped to over-damped oscillations as the ratio $\gamma / g_0$ increases above the critical value $(\gamma / g_0)_c=8$. This can be explained from the behavior of the four eigenvalues $\lambda$ of the matrix in the linear system of Eq.(18-21) versus $\gamma / g_0$ [Figs.2(b) and (c)]. In particular, an exceptional point occurs at $\gamma / g_0=(\gamma / g_0)_c=8$, with all eigenvalues being real for $\gamma / g_0>(\gamma / g_0)_c$. The over-damped relaxation dynamics can be described in the large dephasing limit $\gamma / g_0 \gg 1$ by a reduced set of equations for the populations $\rho_{\epsilon,\epsilon}(t)$ and $\rho_{0,0}(t)$. In fact, in this regime the coherences $\rho_{\epsilon.0}(t)=\rho_{0,\epsilon}^*(t)$ are strongly damped and can be adiabatically eliminated from the dynamics. From Eqs.(20) and (21) one obtains  
\begin{equation}
\rho_{0, \epsilon}(t) \simeq \frac{2 i g_0}{\gamma} \left( \rho_{0,0}-\rho_{\epsilon,\epsilon} \right) , \;\;  
\rho_{ \epsilon,0}(t) \simeq - \frac{2 i g_0}{\gamma} \left( \rho_{0,0}-\rho_{\epsilon,\epsilon} \right).
\end{equation}  
Substitution of Eq.(22) into Eqs.(18) and (19) yields the following rate equations for the populations $\rho_{\epsilon,\epsilon}(t)$ and $\rho_{0,0}(t)$
\begin{eqnarray}
\frac{d \rho_{\epsilon,\epsilon}}{dt}  & = & \mathcal{R} (\rho_{00}-\rho_{\epsilon,\epsilon})  \\
\frac{d \rho_{0,0}}{dt}  & = & - \mathcal{R} (\rho_{00}-\rho_{\epsilon,\epsilon})  
\end{eqnarray}
with the transition rate $\mathcal{R}=4 g_0^2/\gamma$. 

 \begin{figure}
\includegraphics[width=8.5 cm]{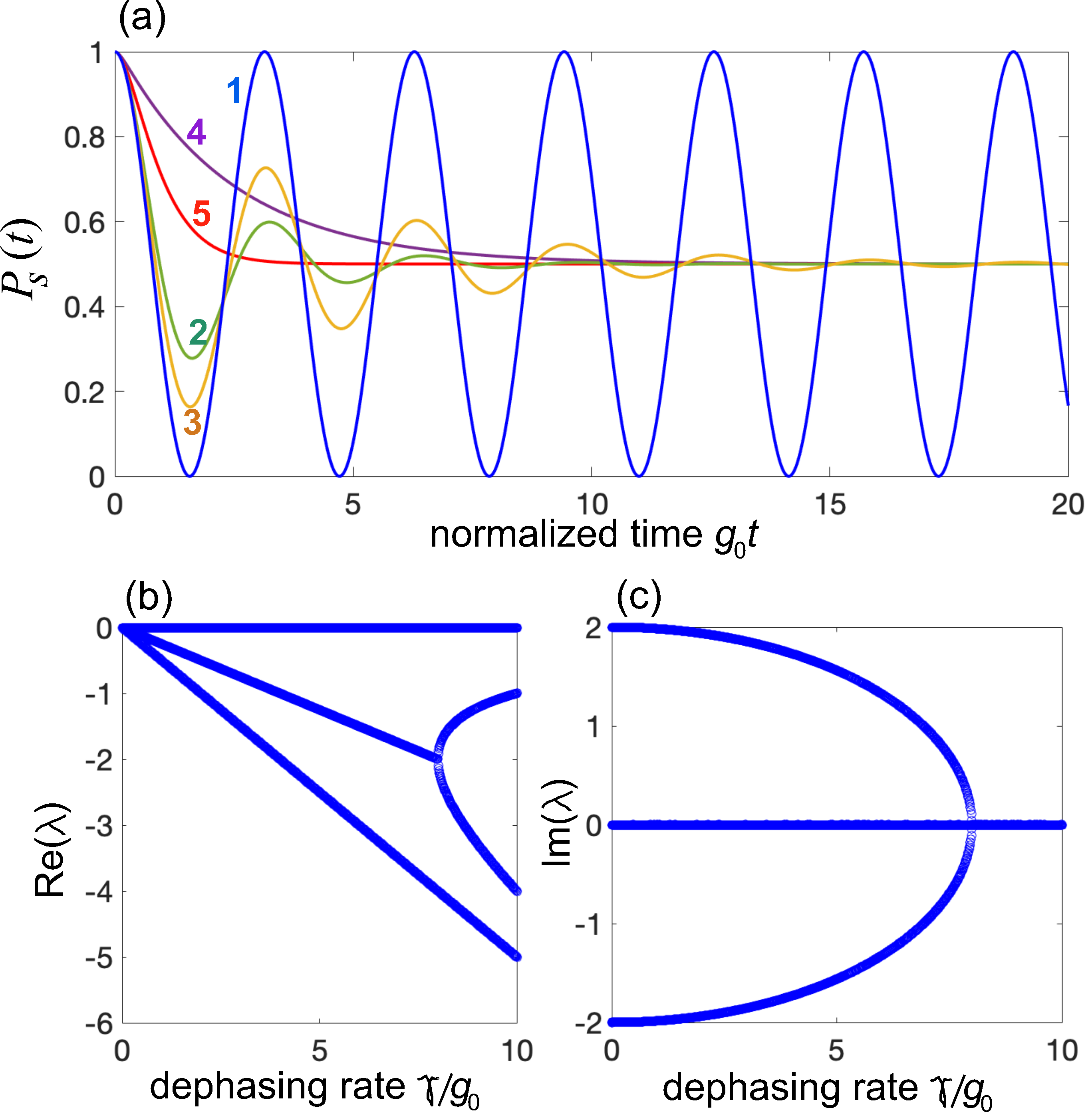}
\caption{(a) Damped vacuum Rabi oscillations induced by dephasing effects for $J=0$, $\omega_0=\omega_c$.and for a few increasing values of the dephasing rate $\gamma /g_0$: Curve 1: $\gamma /g_0=0$; curve 2: $\gamma /g_0=1$; curve 3: $\gamma /g_0=2$; curve 4: $\gamma /g_0=8$; curve 5: $\gamma /g_0=20$. (b,c) Behavior of the eigenvalues $\lambda$ versus the dephasing rate $\gamma / g_0$ (real and imaginary parts) of the relaxation matrix entering in Eqs.(18-21). An exceptional point, corresponding to the coalescence of two of the four eigenvalues and corresponding eigenvectors, is observed at the critical value $(\gamma/g_0)_c=8$. }
\end{figure}

 \begin{figure}
\includegraphics[width=8.5 cm]{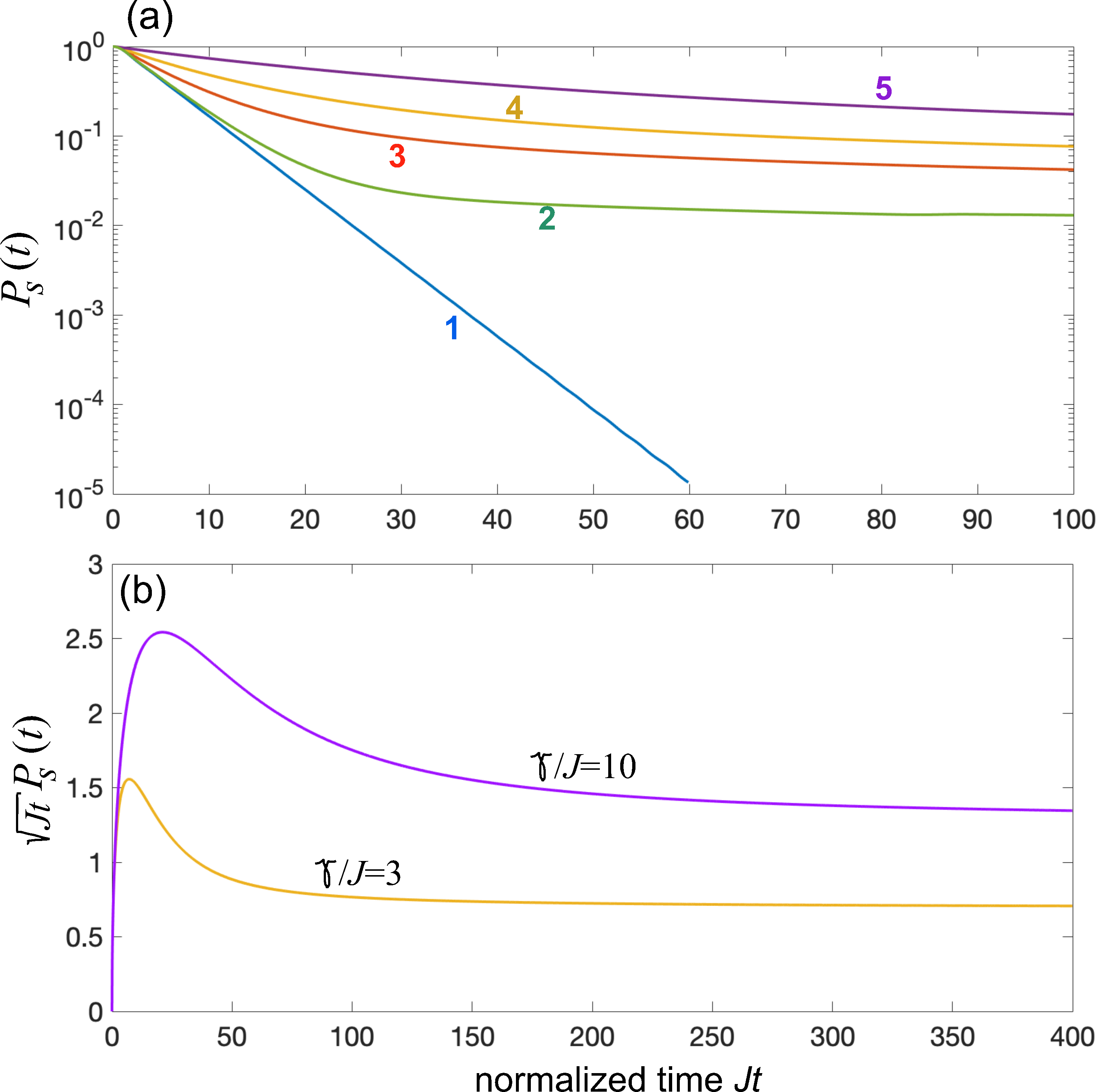}
\caption{(a) Numerically-computed survival probability $P_s(t)$, plotted on a vertical logarithmic scale as a function of normalized time $Jt$ for $g_0/J=0.3$, $\omega_c=\omega_0$  and for several increasing values of the dephasing rate $\gamma /J$. Curve 1: $\gamma /J=0$; curve 2: $\gamma /J=0.1$; curve 3: $\gamma /J=1$; curve 4: $\gamma /J=3$; curve 5: $\gamma /J=10$. (b) Survival probability $P_s(t)$ multiplied by $\sqrt{Jt}$, plotted on a linear scale versus normalized time $Jt$, shown for the strong dephasing regime ($\gamma /J=3$ and $\gamma/J=10$). The curves plateau to a non-zero stationary value, indicating a power-law decay of the form $P_s(t)  \sim 1/  \sqrt{Jt}$.}
\end{figure}

\subsection{Spontaneous emission decay under dephasing}
The spontaneous emission decay is strongly affected by dephasing of the photon modes, leading to pronounced deviations from the exponential decay law as the dephasing rate increases. In particular, a power-law decay emerges at relatively early times, when the survival probability remains appreciable. As an illustrative example, Fig.~3(a) shows numerically computed decay curves for the survival probability \( P_s(t) \), calculated for \( \omega_c = \omega_0 \), \( g_0/J = 0.3 \), and several increasing values of the dephasing rate \( \gamma/J \). The figure clearly demonstrates strong deviations from exponential behavior, with a significant slowdown of relaxation that sets in during the early stages of decay once the dephasing rate becomes comparable to or exceeds \( J \). As shown in Fig.~3(b), in the strong dephasing regime the relaxation follows a power-law trend, and the survival probability decays as $
P_s(t) \sim \frac{1}{\sqrt{Jt}}$. Such a pronounced power-law decay, which emerges already at early times -- contrary to the predominantly exponential behavior observed in the dephasing-free regime shown in Fig.~1(b) -- is driven by photon diffusion in the dynamically disordered coupled-resonator optical waveguide array, rather than by spectral edge effects. In fact, in the strong dephasing limit \( \gamma / J \gg 1 \), the coherences \( \rho_{n,m}(t) \) ($ n \neq m$) and \( \rho_{\epsilon,n}(t) \), which appear in the quantum master equation [Eqs.~(10)--(12)], become small and can be adiabatically eliminated from the dynamics using a procedure analogous to that employed in the analysis of damped vacuum Rabi oscillations (Sec.~III.B; see also Ref.~\cite{Longhi1,Longhi2}). This leads to the following rate equations for the atomic population \( \rho_{\epsilon,\epsilon}(t) \) and the photon populations \( \rho_{n,n}(t) \)
\begin{eqnarray}
\frac{d \rho_{\epsilon,\epsilon}}{dt} & = & \mathcal{R} \left( \rho_{0,0}-\rho_{\epsilon,\epsilon} \right) \\
\frac{d \rho_{0,0}}{dt} & = &- \mathcal{R} \left( \rho_{0,0}-\rho_{\epsilon,\epsilon} )+ \mathcal{Q}(\rho_{1,1}-\rho_{0,0} \right) \\
\frac{d \rho_{n,n}}{dt} & = &\mathcal{Q} \left( \rho_{n+1,n+1}+\rho_{n-1,n-1}-2 \rho_{n,n} \right)  \;\; ( n \geq 1)
\end{eqnarray}
where the rates $\mathcal{R}$ and $\mathcal{Q}$ are given by
\begin{equation}
\mathcal{R}= \frac{4 g_0^2}{\gamma} \; ,\;\;\; \mathcal{Q}=\frac{2J^2}{\gamma}.
\end{equation}
Clearly, the rate equations (25-27) reduce to Eqs.(23,24), describing damped vacuum Rabi oscillations, in the $J=0$ limit.  For $J \neq 0$,  the spontaneous emission process of the excited atom is basically described by a classical continuous-time random walk on a lattice, as schematically illustrated in Fig.4(a). The initial excitation, i.e. walker position, is at the left edge of the lattice, and stochastic hopping with probability rates $\mathcal{R}$ and $\mathcal{Q}$ occurs. From general results of classical random walks on a lattice (see e.g.  Sec.6 of \cite{diffusion}), it is known that the spreading along the lattice is diffusive, which makes it likely a power-law decay $ \sim 1/\sqrt{\mathcal{Q}t}$ of the atomic population $\rho_{\epsilon,\epsilon}(t)$, i.e. of the survival probability of the walker to remain in its initial position, as it is observed in the numerical simulations. This power-law decay can be rigorously proven by considering the exact solution to the rate equations (25-27) corresponding to the initial condition $\rho_{\epsilon,\epsilon}(0)=1$ and $\rho_{n,n}(0)=0$ ($n=0,1,2,3,...$). The exact solution can be obtained using either the resolvent or Fano diagonalization methods (see e.g. \cite{Fano}). As shown in the Supporting Information, the solution reads
\begin{equation}
P_s(t)=\rho_{\epsilon,\epsilon}(t)=\int_0^{\pi} d \omega G(\omega) \exp[E(\omega) t]
\end{equation}
where we have set
\begin{eqnarray}
E(\omega) & = & -2 \mathcal{Q} \left( 1-\cos \omega \right ) \\
G(\omega) & = & \frac{ r^2 (1+ \cos \omega)}{\pi  \left[ r^2+4(r-1)(1-\cos \omega)  (r-1+\cos \omega) \right] } 
\end{eqnarray}
and 
\begin{equation}
r \equiv \mathcal{R}/ \mathcal{Q}.
\end{equation}
Note that $E(\omega) \leq 0$ and $E(\omega) \rightarrow 0^-$ as $\omega \rightarrow 0^+$. Therefore, in the large $t$ limit the dominant contribution to the integral on the right hand side of Eq.(29) comes
from the values of $\omega$ close to $\omega=0^+$. By letting $G(\omega) \simeq G(0) =(2 / \pi) \neq 0$ and $E(\omega) \simeq -\mathcal{Q}\omega^2$ under the sign of integral on the right hand side of Eq.(29), for $\mathcal{Q} t \gg 1$ one can write
\begin{equation}
\rho_{\epsilon,\epsilon}(t) \simeq G(0) \int_{0}^{\infty} d \omega \exp(- \mathcal{Q} \omega^2t)  = \sqrt{\frac{1} {\pi \mathcal{Q}t }}.
\end{equation}
The analytical predictions based on the classical rate equations (random walk) in the strong dephasing regime $J / \gamma \ll 1$ reproduce with excellent accuracy the full numerical solution to the quantum master equation, as shown in Fig.4(b) in an illustrative example.

 \begin{figure}
\includegraphics[width=8.5 cm]{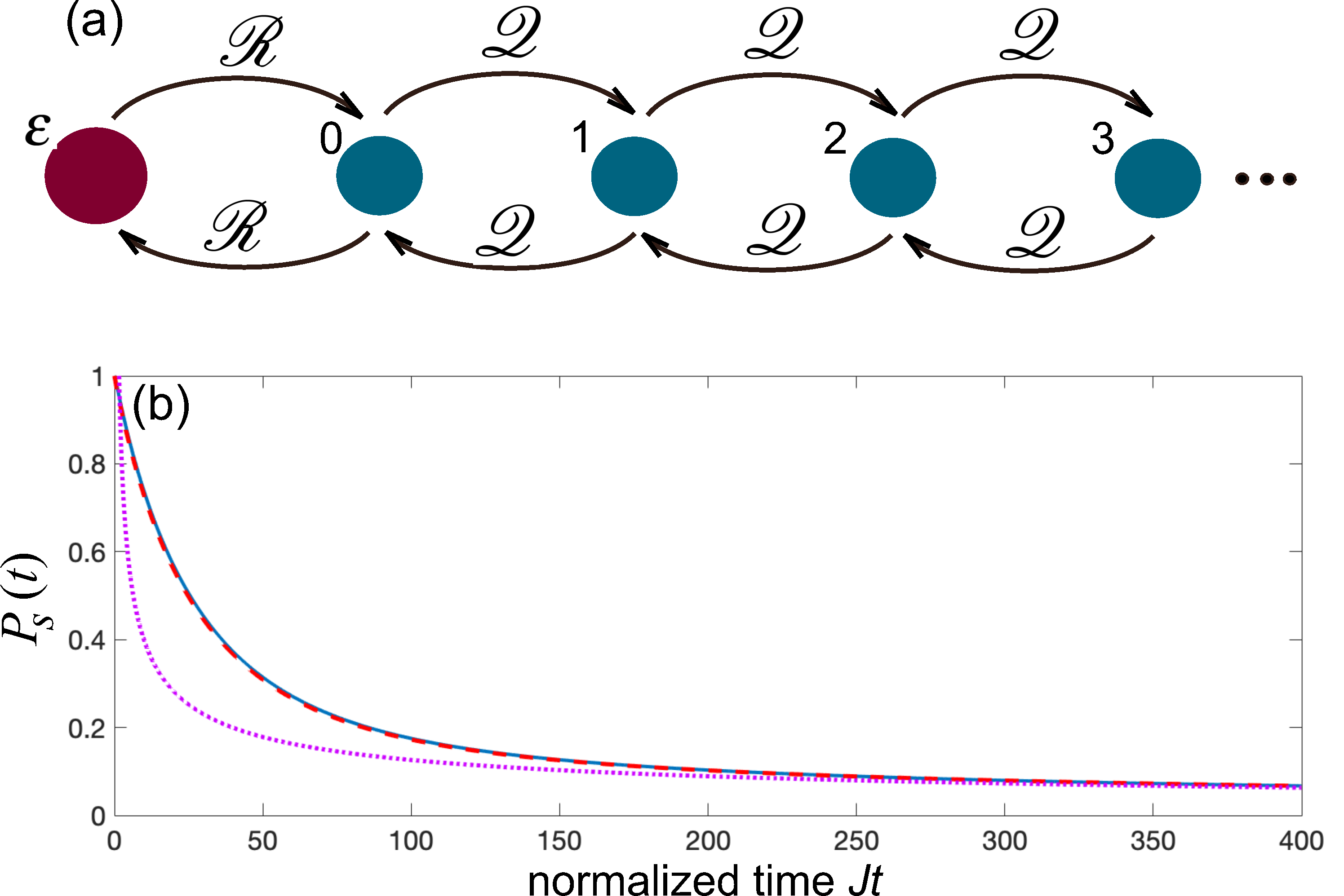}
\caption{(a) Schematic of a classical random walk on a semi-infinite line describing the spontaneous emission process of the two-level atom (at the left edge of the line) under strong dephasing of the photon modes. The hopping rates $\mathcal{R}$ and $\mathcal{Q}$ are given by Eq.(28). (b) Survival probability $P_s(t)$ versus normalized time $Jt$ in the strong dephasing regime ($g_0/J=0.3$ and 
$\gamma /J=10$). The solid grey curve corresponds to the numerical results obtained by solving the quantum master equation [Eqs.(10-12)], the dashed red curve, almost overlapped with the solid one, is the theoretical prediction based on the classical random walk approximation [Eq.(29)], and the purple dotted curve is the asymptotic power-law decay curve given by Eq.(33).}
\end{figure}

\subsection{Discussion}

Throughout this work, we have focused on spontaneous emission in a specific system geometry: a semi-infinite CROW, with the two-level atom embedded in the edge resonator. This configuration offers a clear setting to compare long-time power-law tails arising from spectral edge effects and from dynamical dephasing of the photon field. In the absence of dephasing, the long-time power-law decay typically observed originates from spectral edge effects and is highly sensitive to the global geometry of the resonator network. Variations in the waveguide structure -- such as semi-infinite vs. infinite extent, boundary conditions, or local disorder -- can significantly alter or even suppress the spontaneous emission decay.
For instance, in an infinite linear CROW, spontaneous emission can be inhibited due to the formation of atom-photon bound states, which trap a portion of the excitation and prevent complete decay. Likewise, introducing static disorder in the resonator frequencies leads to Anderson localization of photonic modes, giving rise to Rabi-like population oscillations and incomplete relaxation dynamics, as observed in previous studies (see e.g. \cite{D23}). 
By contrast, the power-law decay behavior induced by dynamical dephasing of the photonic modes, as revealed in this work, arises from a different mechanism -- namely, photon diffusion in a dynamically disordered environment. Crucially, this decoherence-driven decay is robust and largely insensitive to the underlying photonic network topology or static disorder. The presence of dephasing effectively washes out the coherent interference patterns and localized modes responsible for bound-state formation and localization effects, reinstating a universal power-law decay regime even in configurations where conventional decay is otherwise strongly suppressed. \textcolor{black}{Importantly, the decoherence-induced power-law decay predicted here is fundamentally distinct from the conventional long-time algebraic decay associated with the continuum threshold: it originates from photon diffusion due to dynamic dephasing rather than spectral edge effects, occurs on experimentally accessible timescales, and strictly requires time-dependent disorder, as static (Anderson) disorder in one-dimensional systems cannot induce diffusion.}
\textcolor{black}{A physical picture for the time-dependent disorder considered here can be provided in terms of random, uncorrelated fluctuations of the resonator frequencies or cavity couplings, which can be engineered in photonic platforms by controlled modulation of the cavity parameters,  in superconducting circuits by applying fast, stochastic flux noise, or in cold-atom systems via controlled fluctuations of external fields or laser intensities. Unlike static disorder, which produces localized modes, these temporal fluctuations induce a dephasing process that allows photons to diffuse through the array, giving rise to the observed power-law decay. While some experimental correlations in the noise are inevitable, the predicted effect is robust as long as the correlation time is shorter than the characteristic photon hopping timescale.}
These findings suggest that decoherence, typically regarded as detrimental, can play a constructive role in restoring generic and robust power-law decay dynamics in structured quantum environments. This mechanism is expected to persist across a broad class of CROW architectures, including disordered or topologically nontrivial networks, highlighting the universality and potential experimental relevance of the effect.

\section{Conclusions}
In this work, we have unveiled a novel mechanism for non-exponential spontaneous emission in waveguide quantum electrodynamics, driven by dynamical dephasing in the photonic environment. Contrary to the conventional scenario where power-law decay emerges only at long times due to spectral edge effects, we have shown that introducing dephasing into the photon modes induces a robust power-law decay already at short times. This behavior originates from photon diffusion in the dynamically disordered waveguide, marking a fundamentally different, decoherence-induced pathway to power-law decay.
Our results reveal that decoherence -- typically associated with loss of coherence and detrimental effects-- can instead play a constructive and defining role in shaping quantum emission dynamics. This finding highlights the potential of engineered decoherence as a tool to control and probe non-trivial decay behavior in waveguide QED environments.
The predicted power-law regime is experimentally accessible with current quantum technologies, particularly in integrated photonic systems and in superconducting quantum circuits, which offer a highly versatile platform for simulating waveguide QED and engineering tunable decoherence. These insights may inspire experimental verification and open new avenues for controlling spontaneous emission decay in quantum networks.\\
Future research could explore how decoherence-induced power-law decay manifests in more complex settings, such as topological photonic lattices, non-Hermitian environments, or many-emitter systems exhibiting collective decay \cite{conc1,conc2,conc2b,conc3,conc3b,conc4,conc5}. Investigating the effects of strong light-matter coupling, non-classical noise, and time-dependent dephasing protocols may offer further opportunities to tailor spontaneous emission dynamics and exploit non-Markovian effects in open quantum platforms.



\section*{Supporting Information}
Supporting Information is available from the Wiley Online Library or from
the author.

\section*{Acknowledgements}
The author acknowledges the
Spanish Agencia Estatal de Investigacion (MDM-2017-0711).

\section*{Conflict of Interest}
The author declares no conflict of interest.

\section*{Data Availability Statement}
The data that support the findings of this study are available from the corresponding
author upon reasonable request.





\end{document}